\documentclass[aps,twocolumn,floats,superscriptaddress]{revtex4}
\usepackage{amsmath}
\usepackage{graphicx}

\def\moy#1{\left\langle #1 \right\rangle}

\def\figx#1#2{\includegraphics[width=#1]{#2}}

\newcommand{\rarrow}{\rightarrow }

\def\etal{{\it et al.\/}}

\begin{document}

\title{Is the quantum dot at large bias a weak-coupling problem?}

\author{P. Coleman}
\affiliation{Center for Materials Theory,
Department of Physics and Astronomy, Rutgers University, Piscataway, NJ
08854, USA}

\author{C. Hooley}
\affiliation{Center for Materials Theory,
Department of Physics and Astronomy, Rutgers University, Piscataway, NJ
08854, USA}  
\affiliation{
School of Physics and Astronomy, Birmingham University, Edgbaston,
Birmingham B15 2TT, UK}

\author{O. Parcollet}
\affiliation{Center for Materials Theory,
Department of Physics and Astronomy, Rutgers University, Piscataway, NJ
08854, USA}  

\begin{abstract}
We examine the two-lead Kondo model for a d.c.\ biased 
quantum dot in the Coulomb blockade regime.
From perturbative calculations of the 
magnetic susceptibility,
we show that the problem retains its
strong-coupling nature, even at bias voltages larger than the
equilibrium Kondo temperature.
We give a speculative discussion of the nature of the renormalization group flows 
and the strong-coupling state that emerges at large voltage bias.
\end{abstract}

\maketitle

For over a decade, transport measurements on quantum dot systems have
stimulated 
theoretical interest in the properties of the Anderson model out of
equilibrium \cite{Kastner:1993,Meirav:1996,Ashoori:1996}.  
Early 
predictions \cite{Glazman:1988,Ng:1988,Meir:1992,Meir:1993,Wingreen:1994,Konig:1996} of the Kondo effect in such systems have
recently been experimentally 
verified \cite{GG:1998,Cronenwett:1998,Schmid:1998,vanderWiel:2000}.  In the small voltage regime, such
experiments have produced
impressive agreement with theoretical predictions. Recent interest
\cite{Glazman:2000,Langreth,Schiller:1998} has turned to the question of how the Kondo effect
behaves far from equilibrium, i.e.\ at large voltage bias. 

In the Coulomb blockade regime, one may neglect
charge fluctuations of a quantum dot 
and concentrate on transport processes that operate by
flipping the dot's spin.  In this limit, the physics
is described by a two-lead Kondo model \cite{Avishai:1998,Glazman:2000} (eq.\ (\ref{ham})).
A key property of the Kondo model in
equilibrium is the presence of a ``running
coupling constant'', whereby the antiferromagnetic
coupling between the dot spin and the leads grows
as the energy scale is reduced. This leads to a ``strong
coupling'' low-energy regime where the spin of the dot is
quenched by the lead electrons and the residual 
properties can no longer be obtained from a perturbative expansion. 
A single scale, the ``Kondo temperature'' $T_{K}$, governs the low temperature
properties;
$T_{K} = D \sqrt{g} \, e^{-1/2g}$,
where $g$ is the `bare coupling' 
between the 
spin of the dot and 
the leads and $D \, {\gg} \, T_{K}$ is the electron bandwidth.
Thus the
magnetization at temperature $T$ and magnetic field $B$ is
a universal function
$M = m (T/T_K,B/T_K)$,
where $M$
has a perturbative ``weak coupling'' expansion 
in $g$ only when $T, B \gg T_K$. 
(See Fig.~1(a).) 

In a quantum dot, the formation of 
an Abrikosov-Suhl resonance in the quasiparticle density of states
which accompanies
the Kondo effect 
substantially enhances the linear conductivity
at low temperature, resulting in unitary transmission at absolute
zero \cite{Glazman:1988,Ng:1988}. 

\begin{figure}[ht]
\[
\figx{8.6cm}{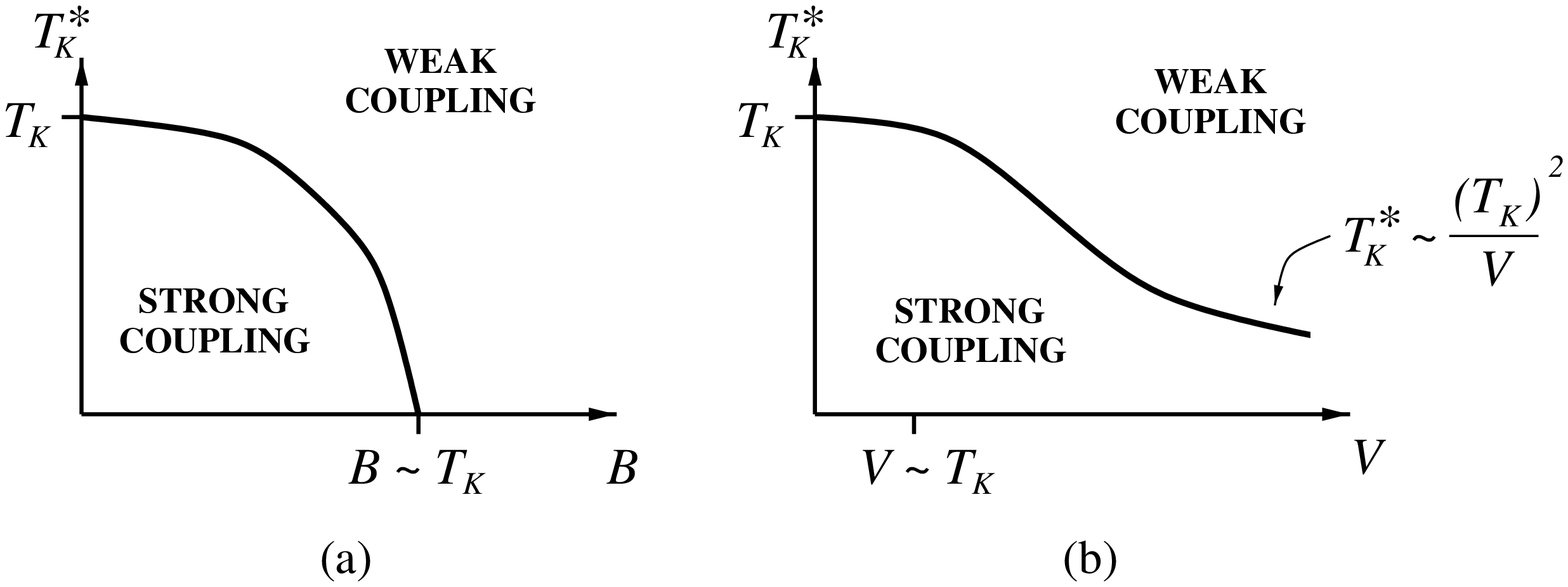}
\]

\vspace*{-4mm}
\caption{\label{FigureT*}
\small
(a) Field dependence of the crossover temperature $T_K^*(B)$ separating 
weak- and strong-coupling regimes of the equilibrium Kondo model.
At absolute zero, for fields larger than the Kondo temperature
the quantum dot re-enters weak coupling. 
(b) Voltage dependence of the 
crossover temperature $T_K^*(V)$, from our results.  Note that $T_K^*$ goes
to zero only in the limit $V \to \infty$, i.e.\ for low enough temperatures, the model
reaches a strong coupling state for all $V$.}
\end{figure}

But what happens to the physics of the quantum dot at voltages 
$V$ comparable to or larger than
the Kondo temperature? In particular, does a 
large voltage act like a large magnetic field or temperature, 
and return the dot to a weak coupling regime where its spin is unquenched?
Theoretical studies of the differential conductivity $G[V]$
do suggest that 
the large voltage physics is governed by weak-coupling perturbation
theory, for \cite{Glazman:2000}
$G = (2e^2/h) F [ eV/T_K ]$
has a perturbative expansion in $g$ for $V \gg T_{K}$.
However, the large bias conductivity
probes electrons that are far from the Fermi surface
and does not in itself shed light on the coupling between
the local moment and the electrons near the Fermi surface.

In this letter, we argue that the quantum dot retains
its strong-coupling character even at voltages $V \gg T_K$.
We use explicit perturbative calculations of the magnetic
susceptibility to show that the large 
bias quantum dot is characterized by a strong-coupling regime with
a voltage-dependent Kondo temperature $T_K^* \propto T_K^2/eV$. 

The Hamiltonian of the two-lead Kondo model is
\begin{eqnarray}\label{ham}
H & = & \sum_{\alpha {\bf k}\sigma} \varepsilon_{\alpha{\bf k}} c^\dag_{\alpha
{\bf k} \sigma} c_{\alpha {\bf k} \sigma} + H_{\rm refl} + H_{\rm trans}; \\
H_{\rm refl} & = & J_{R} 
\sum_{{\bf k},{\bf k}',\sigma,\sigma'}
\left(
c^\dag_{R{\bf k}\sigma} {\vec \sigma}_{\sigma \sigma'}
 c_{R {\bf k}' \sigma'} \right)
\cdot {\vec S} \,\, + \,\, (R \to L), \nonumber \\
H_{\rm trans} & = &
J_{LR} 
\sum_{{\bf k},{\bf k}' , \sigma,\sigma'}
 \left(
c^\dag_{R{\bf k}\sigma} {\vec \sigma}_{\sigma \sigma'} c_{L {\bf k}' \sigma'} \right)
\cdot {\vec S} \,\, + \,\, (R \leftrightarrow L). \nonumber
\end{eqnarray}
Here, $c^\dag_{\alpha {\bf k} \sigma}$ creates an electron in lead
$\alpha \in \left\{L,R\right\}$ with momentum ${\bf k}$ and spin
$\sigma$, and $J_L$, $J_R$ and $J_{LR}=J_{RL}$
are positive (antiferromagnetic) Kondo coupling constants between the
electrons and the dot ($\vec{S}$).  

The first part of $H$ describes 
the electrons in the leads, with energies
$\varepsilon_{\alpha{\bf k}}=\varepsilon_{{\bf k}}-e V_{\alpha}
$, where $V_{\alpha}=\pm V/2$ are the potentials of the left- and right-hand
leads.
$H_{\rm refl}$ describes regular Kondo processes, where an electron from a given
lead is spin-flip scattered back into the {\em same\/} lead; $H_{\rm trans}$
describes `spin-flip cotunnelling', where an electron from one lead is
spin-flip scattered into the {\em other\/} lead.
We introduce the
dimensionless couplings ${\bar J}_L,{\bar J}_R,{\bar J}_{LR}$, defined by
${\bar J}_i \equiv \rho J_i$, where $\rho$ is the density of states at the Fermi levels.
When the model is derived from an
Anderson model via a Schrieffer-Wolff transformation \cite{Schrieffer:1966,
Kaminski:1999}, the coupling
constants obey $|J_{LR}|^{2} = J_{L}J_{R}$, restricting
the scattering to a single channel. Symmetric dots
($J_{L}{=}J_{R}{=}J_{LR}$), are of particular interest, for in
this case the differential conductivity $G$ approaches
the unitary limit ($2e^2/h$) at low
temperatures.

In a magnetic field we replace $H\rarrow H - B M_z$, where 
$
M_z=2S_z +
\sum_{\alpha{\bf k}\sigma}\sigma n_{\alpha{\bf k}\sigma}$ 
 is the
magnetization. The 
impurity susceptibility is
${\chi}_{i} = \frac{\partial}{\partial B} M_z \vert_{B=0} - \chi_o$,
where $\chi_o= 4 \rho $ is the Pauli susceptibility of the leads. 
We expand $\chi _{i}$ perturbatively to order
$J^2$ using the Keldysh method \cite{Rammer:1986}, and representing the spin as
${\vec S} = - {i \over 2} {\vec \eta \times \vec \eta}$,
where the elements of ${\vec \eta}$ are Majorana fermions.
In the large-bandwidth limit we find that
\begin{multline}\label{suscept}
\chi_{i} = \frac{1}{T}
\left\{
 1 - \frac{1}{2}(\bar J_{R} +\bar J_{L}) -
\left(
   \bar J_L^2 + \bar J_R^2
\right)
\ln \left(
       \frac{D}{2\pi T e^{1+\gamma}}
    \right) 
\right. 
\\
\left.
- 2|\bar J_{LR}|^2 \left[ \ln
 \left(
\frac{D}{ 2\pi T e^{1+\gamma}} 
\right)
- \phi  \left( \frac{V}{T} \right)
\right] 
\right\}.
\end{multline}
where $\gamma=0.577\dots$ is the Euler constant.  The
crossover function $\phi(x)$, in terms of digamma
functions $\psi(z)$, is
\begin{equation}\label{DefPhi}
\phi (x)  =
\hbox{Re} \int_{-\infty }^\infty 
\frac{dy}{4 \cosh^2 \bigl(\frac{y}{2}\bigr)}\left[
\tilde{\psi } (y+x)-\tilde{\psi}(y) \right]
\end{equation}
where $\tilde{\psi}(x)=\psi ( \frac{1}{2} + i \frac{x}{2\pi} )$.

The second-order terms in (\ref{suscept}) describe the leading
logarithmic enhancement of the Kondo coupling. 
Terms of order $J_{LR}^2$ involve inter-lead
processes and, as expected, the 
logarithmic divergence in these terms 
is cut by the voltage.  (To
see this, note that $-\phi (V/T)\sim -\ln (V/T)$ for $V \gg T$ which 
cancels the logarithmic temperature divergence.) 
By contrast, the intra-lead terms of order $J_R^2$ and $J_L^2$
are completely unaffected by the voltage $V$, which guarantees that the
leading logarithmic divergence survives
at {\em arbitrarily high voltage}.  This is easily seen in the large-$V$
form of the susceptibility,
\begin{multline}\label{largeVsuscept}
\chi_{i} = \frac{1}{T}
\left\{
 1 - \frac{1}{2}(\bar J_{R} +\bar J_{L})  -
 \left ( 
       \bar J_L^2 + \bar J_R^2
 \right)
\ln \left(
        \frac{D}{2\pi T e^{1+\gamma}}
    \right)
\right.
\\
\left.
- 2|\bar J_{LR}|^2  \ln \left( \frac{D}{V} \right)
\right\}, \qquad \quad V \gg T.
\end{multline}
The survival of these leading logarithms
is a signature that the intra-lead Kondo effect 
continues unabated at temperatures smaller than the voltage. 

Let us now be more precise by defining a {\em generalized crossover
temperature\/} $T_K^*$ as the temperature below which the perturbation expansion
(\ref{suscept}) breaks down, i.e.\ the temperature at
which the $O(J)$ and $O(J^2)$ terms become equal in magnitude.
This procedure yields an implicit formula for $T^{*}_{K} (V)$
\begin{equation}\label{EquationForT*}
T^*_K (V) = T_{K}e^{-\frac{1}{2}\phi (V/T_K^* (V))}
\end{equation}
where $T_{K} = D e^{-1/2g}$ is
the leading approximation to the Kondo temperature \cite{note}.
We plot $T_K^*(V)$ in Fig.~1(b).

Note that $T_K^*$ goes to
zero only when $V$ goes to infinity.  From the behavior of $\phi$ at large
argument, we can extract the large voltage behavior
\begin{equation}\label{LargeVoltageBehaviour}
T^*_K = \frac{(T_{K})^{2}}{V}.
\end{equation}
This means that the model always enters a strong coupling regime at
low temperature, even for $V \gg T_{K}$.

These perturbative results for $\chi_i$ may be
interpreted within a simple renormalization picture.
Let us stress, however, that there is not yet a general theory of
renormalization out of equilibrium (though see \cite{Schoeller}).
In particular, it has not been established that there exists an
appropriate family of
effective Hamiltonians through which one passes as the high-energy
cutoff scale is varied.  Furthermore, in the presence of a voltage,
long-time processes cannot be unambiguously connected with
low-energy ones.
Hence the following discussion
is somewhat speculative, and steps beyond the perturbative calculation
that led to (\ref{EquationForT*}).

In the spirit of a ``poor man's scaling'' approach \cite{poor},
we consider varying a cutoff energy scale
$\Lambda$ in the problem, and studying the variation of the effective
coupling constants $J_L, J_R, J_{LR}$ as functions of this variable.
The essential scaling behavior, as usual, can be read off from the
perturbative expression (\ref{suscept}).  We may identify two scaling
regimes:

(1) One-channel scaling, $\Lambda \gg V$. In this regime
the effect of the voltage is not felt in any channel, so the 
dot behaves as an equilibrium single-channel Kondo model. 
The single coupling constant $g \equiv {\bar J}_L + {\bar J}_R$ rescales
according to 
\begin{equation}
\frac{d g}{d\ln \Lambda}= -2 g^{2} + O ( g^{3})
\end{equation}
so to leading order in the scaling, the coupling constant 
at scale $\Lambda=V$ is given by
$\frac{1}{g^{*}} = \frac{1}{g} - 2\ln (D/V)$. In this regime,
the ratio $J_{LR}/g$ is preserved.

(2) Two-channel regime $V \gg \Lambda \gg T^*_K(V)$. 
Around $\Lambda\sim V$ the logarithmic
renormalization of inter-lead processes comes to a halt, freezing 
$J_{LR}$ at its value $J_{LR}(V)$ and prompting 
a crossover to a two-channel scaling regime in which intra-lead
processes independently renormalize $J_R$ and $J_L$. This scenario 
is reminiscent of one proposed by Wen \cite{Wen:1998}:\ that the 
effect of the voltage is to destroy coherence {\em between\/} the
two leads, suppressing the $J_{LR}$ terms, and leaving a two-channel Kondo
Hamiltonian. We shall restrict our attention to 
symmetric dots, so that $J_{L,R}=\bar{J} (\Lambda)$ scale together 
according to 
\begin{equation}
\frac{d {\bar J}}{d \ln \Lambda}= -2{{\bar J} \,}^{2} + O (
{{\bar J} \,}^{3}).
\end{equation}
At voltage $V$, ${\bar J}_{R,L}= g^{*}/2$, so to 
leading order, the coupling constants are given by
\begin{equation}\label{rescale}
\frac{1}{{\bar J} (\Lambda)}= \frac{2}{{g^{*}}}  -
2 \ln (V/\Lambda ).
\end{equation}
Scaling continues until these coupling 
constants become of order unity. Setting the left-hand side of
(\ref{rescale}) to one, we recover the same crossover scale  obtained
by comparing terms in the susceptibility:
$
T^*_K\sim  V e^{-1/{g^*}}= (T_{K})^{2}/V.
$

\begin{figure}[ht]
\[
\figx{8cm}{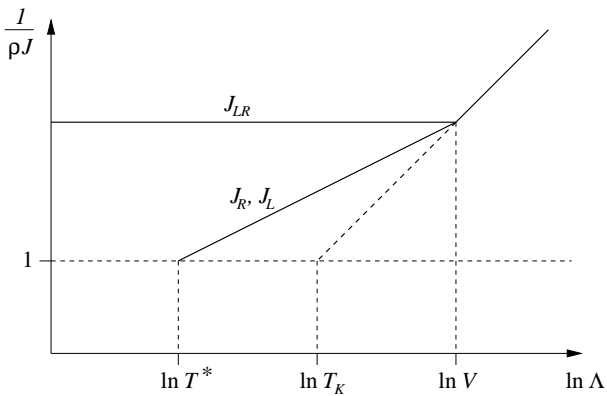}
\]

\vspace*{-4mm}
\caption{\label{flows}
\small
Schematic renormalization flows for this model, for the case
$V>T_K>T$.
All couplings flow together down to scales of order $V$, after which the flow of $J_{LR}$ is halted at
its value $J_{LR}(V)$.  The couplings $J_L$ and $J_R$ continue to flow to strong coupling
values, but more slowly, yielding a generalized crossover temperature
$T_K^*(V)\sim T_{K}^{2}/V$.
}
\end{figure}

These flows are depicted in Fig.~\ref{flows}.

Thus we conclude that, for $V>T_K$, the scaling is of a two-channel
character by the time strong coupling is reached.  This suggests
that an appropriate low-temperature description may be
in terms of the infra-red fixed point of the two-channel
Kondo model.  However, it is known that this point is unstable to
perturbations of the form $J_{LR}$, so how can this be?  We shall show
that the presence of a finite voltage renders the $J_{LR}$ perturbations
irrelevant, making the two-channel infra-red fixed point a
likely candidate for the $V>T_K$ state of the model.

To demonstrate this,
let us suppose that $H_{0}$ is the Hamiltonian where 
$J_{LR}=0$, and consider the effects of
introducing a small left-right coupling $\lambda$.
For convenience, let us perform a gauge transformation on the
lead-electron operators,
$c_{\alpha {\bf k} \sigma} \to e^{i V_\alpha t} c_{\alpha {\bf k}
\sigma}$.
This removes the chemical potentials
from the bare energies of the lead electrons, with the compensating
effect of introducing a factor $e^{\pm iVt}$ into the $J_{LR}$ terms.
Therefore, the perturbation we wish to consider is given by
$
H_{I} (t)=\lambda[ O^{\dagger} e^{{-iVt} } + O e^{{iVt}}],
$
where 
$O =c^{\dagger}_{L}\vec{\sigma }c_{R}\cdot \vec{S} $.

From the conformal field theory of the two-channel
Kondo model \cite{Ludwig-Sengupta},
we know the scaling dimension of the inter-channel
operator, which scales according to 
$\moy{ 
O (t)O^{\dagger } (0)
}
\sim \frac{1}{t}$
at long times. Any perturbation coupling statically to $O$ is
relevant, but we are interested in a coupling 
to the oscillatory operator $O e^{iVt}$.  If we
calculate the expectation value of  $H_{I}$, we
obtain
\begin{equation}\label{core}
\langle H_{I} \rangle = 2\lambda^{2}\,{\rm Im} \!\! \int\limits_{0}^{\infty} \!\! dt
\moy{[O (t),O^{\dagger } (0)]} e^{{-iVt}} \sim \ln (\frac{\tilde{\Lambda}}{V}),
\end{equation}
showing that the perturbation is finite in a non-zero voltage, but becomes
relevant, as expected, in the $V=0$ case.
This suggests that the finite voltage protects the two-channel
fixed point against the damaging effects of inter-lead coupling.
We expect that these arguments can be examined in more detail using
a bosonized formulation of the two-channel Kondo model \cite{Schiller:1998}.

We may go one small step beyond this point, and examine briefly the
nature of the current fluctuations in the leads at this fixed
point. The current operator is given by
$I(t) = - i (e/\hbar) \lambda[ O^{\dagger }e^{-iVt} - {\rm h.c.}]$.
From the scaling dimension of the operators $O$ and $O^{\dagger }$ we
can see that the current-current correlator decays as $e^{{-iVt}}/t$,
so that we expect that for small $\lambda$
\begin{equation}
C(\omega) \equiv \left\langle \left\{ I(\omega),I(-\omega) \right\}
\right\rangle \sim \ln (\vert \omega\vert -V),
\end{equation}
i.e.\ $C(\omega)$
will develop a weak logarithmic divergence at frequencies
$\omega\sim eV/\hbar$.  A similar divergence (resulting from the
corresponding $1/t$ decay of the spin correlators) is responsible
for the logarithmic divergence of the spin susceptibility 
in a two channel Kondo model. 

This discussion leads us to propose the crossover diagram presented in
Fig.~3.
\begin{figure}[ht]
\[
\figx{7.5cm}{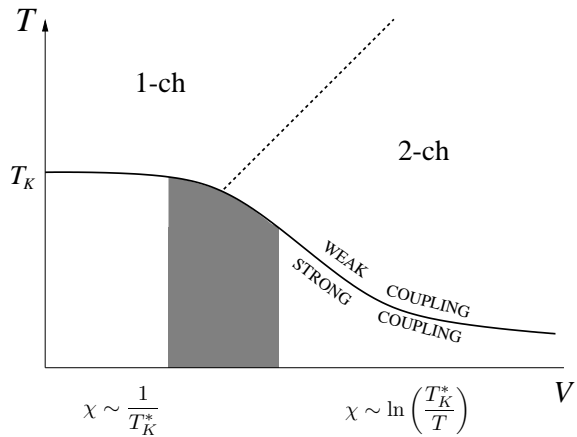}
\]

\vspace*{-4mm}
\caption{\label{phase}
\small
Schematic crossover diagram for the two-lead Kondo model.  
The dashed line separates
the regions $V<T$ and $V>T$ corresponding to one- and two-channel
scaling behavior respectively.
The solid line represents the crossover from the weak- to the strong-coupling region, estimated from
perturbative calculations.}
\end{figure}
If indeed $T^*_K $ is the only
scale in the problem, the susceptibility and
frequency-dependent conductivity in this system are
expected to display the following logarithmic forms
at low temperature:
\begin{eqnarray}\label{guess}
G (\omega) & \sim & \frac{e^{2}}{\hbar }\ln
\left[\frac{T^*_K}{\hbox{max} [|\omega-V|, T]} \right], \\
\chi (T,V) & \sim & \frac{1}{T^*_K}\ln\left[\frac{T^*_K}{T} \right].
\end{eqnarray}
If these tentative forms hold true, then the spin susceptibility
must go from a finite value at low voltages to a divergent form
at high voltages, suggesting the possibility of a quantum critical
point separating the low and high voltage regimes of the quantum
dot. 

We end by adding a brief remark on the issue of decoherence.
A common view is that large voltages decohere the spin fluctuations of
the quantum dot, introducing a decoherence time 
given by 
$
\tau^{-1} \approx (\rho J)^2 V.
$  (See, for example, eq.~(43) of \cite{Wingreen:1994} and
p.~387 of \cite{Kaminski:1999}.)
If this decoherence is to destroy the Kondo effect, it must cut the
logarithms in expressions like (\ref{suscept}) before strong coupling
is achieved.  But 
in order for terms of the type 
$
\ln \left( D/ \mbox{max}\, (\tau^{-1},T) \right)
$
to appear in the high temperature expansion, the perturbation theory
would have to break down at a temperature larger than $T^*_K$, for
which there is no evidence in the low-order perturbation theory.
Unless such a breakdown occurs at higher orders,
we conclude that $\tau^{-1} < T_K^*$.

To conclude, we have presented perturbative arguments which suggest
that the two-lead Kondo model, representing a quantum dot deep in the
Coulomb blockade regime, retains its strong-coupling nature even for
voltages $V>T_K$.  To reach this conclusion, we have used a
perturbative calculation of the magnetic susceptibility to read off
the leading renormalization behavior.  We concluded that $J_{LR}$
ceases to renormalize at scales of order $V$, while $J_L$ and $J_R$
continue to renormalize, though more slowly, towards strong coupling.
This suggests a picture in which the model scales to a two-channel
infra-red fixed point.  To test this picture, we performed a partial
analysis of the stability of this point
with respect to small perturbations of the form $J_{LR}$.  At finite
voltages, the point was found to be stable to such perturbations,
although the r{\^o}le of dangerous irrelevant operators may
warrant further inspection.
The voltage dependence of $T_K^*$, and the
associated crossover to two-channel Kondo behavior, imply that the
voltage, rather than simply providing a frequency shift, qualitatively changes
the structure of the model's excitation spectrum.

We would like to thank C.\ Bolech,
R.\ Aguado, D.\ Langreth, A.\ Kaminski, and L.\ Glazman for useful
discussions.
We note that related ideas are being  pursued independently by
Carlos Bolech and Natan Andrei using a Keldysh effective action.
This work was supported by DOE grant DE-FG02-00ER45790 (PC),
NSF grant DMR 99-76665 (OP), EPSRC fellowship GR/M70476 (CH),
and the Lindemann Trust Foundation (CH).

$\mathstrut$

\end{document}